\documentclass{aa} 
\usepackage{epsf}
\usepackage{graphicx}

\newcommand{\cts}{$\rm {cts\,s}^{-1}$}

\newcommand{\RX}{RX~J1313.2--3259}

\newcommand{\fluxunit}{$10^{-16}$\,erg\,cm$^{-2}$s$^{-1}$\AA$^{-1}$}
\newcommand{\Ion}[2]{#1{\,\scriptsize #2}}
\begin{document}

\thesaurus{06(08.14.2, 08.13.1, 08.09.2 \RX{}, 08.02.1, 13.25.5 ,02.01.2)}
\title{RX~J1313.2--3259, a long-period Polar discovered with ROSAT
\thanks{Based in part on observations collected at the European Southern
Observatory, La Silla, Chile with the ESO/MPI 2.2m telescope in MPI time
and with various telescopes in ESO time (ESO Nos. 50.6-017, 50.6-021,
54.D-0698, 55.D-0383, 56.D-0552, 56.D-0561, 61.D-0415).
}}
\author{H.-C. Thomas\inst{1},
K. Beuermann\inst{2,3},
V. Burwitz\inst{3,2},
K. Reinsch\inst{2},
\and A.D. Schwope\inst{4},
}

\offprints{hcthomas@mpa-garching.mpg.de}

\institute{MPI f\"ur Astrophysik, Karl-Schwarzschild-Str. 1, D-85740 Garching, Germany
\and Universit\"ats-Sternwarte, Geismarlandstr. 11, D-37083 G\"ottingen, Germany
\and MPI f\"ur extraterrestrische Physik, Giessenbachstr. 6, D-85740 Garching, Germany
\and  Astrophysikalisches Institut Potsdam, An der Sternwarte 16, D-14482 Potsdam, Germany
}
\date{Received 7 September 1999 / Accepted 5 November 1999}

\authorrunning{H.-C. Thomas et al.}
\titlerunning{~RX~J1313.2--3259, a long-period Polar}

\maketitle 


\begin{abstract}

We report observations of a new AM Herculis binary identified
as the optical counterpart of the X-ray source \RX{},
detected during the ROSAT All-Sky Survey (RASS). It has an orbital period
of 251~min and is strongly modulated at optical wavelengths. The
long-term behavior is characterized by a pronounced variation in X-rays
between the RASS and two subsequent pointings (decrease
by a factor 40 in count rate) and by moderate changes in the optical brightness (up to a
factor 5). The X-ray spectrum is dominated by a soft quasi-blackbody component,
with a smaller contribution from thermal bremsstrahlung. Measurements of
high circular polarization confirm its classification as a Polar with a
magnetic field strength of 56~MG. The average visual magnitude of \RX{} is
$V \simeq 16^{\rm m}$, for its distance we get $\simeq$~200~pc.


\end{abstract} 

\section{Introduction} 

\begin{figure}[t]
 \centerline{
 \epsfxsize=8.8cm
 \epsfbox[70 290 527 758]{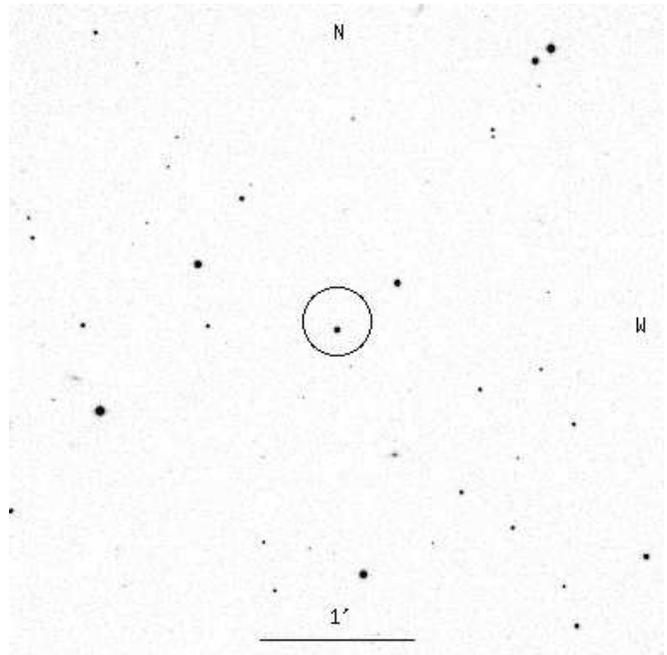}}
 \caption{V-band CCD image of \RX{}. The X-ray error circle from the
 RASS refers to the 90\% confidence level. The position of the optical
 counterpart is $\alpha_{2000}=13^{\rm h} 13^{\rm m} 17\fs 1$, 
 $\delta_{2000}=-32^\circ 59' 12\arcsec$, the average visual magnitude $V \simeq 16^m$.}
\end{figure}

\begin {table*}
\caption[]{ List of observations. Columns denote: (1) and (2) the time of the observation,
(3) and (4) the telescope and instrument used, (5) the type of observation (spec.:
spectrophotometry, point.: X-ray pointing, phot.: photometry, sp.pol.: spectropolarimetry),
(6) the spectral range or the filters used, (7) the FWHM resolution (spectroscopy only),
(8) the number of spectra or V filter data obtained, (9) the time span for the observation,
and (10) the exposure time for the individual measurements.
}
\begin {flushleft}
\begin {tabular}{rrlllcrrr@{\hspace{1mm}}lr@{\hspace{1mm}}l}
\noalign{\smallskip}
\hline
\noalign{\smallskip}
\multicolumn{1}{c}{(1)} & \multicolumn{1}{c}{(2)} & \multicolumn{1}{c}{(3)} & 
\multicolumn{1}{c}{(4)} & \multicolumn{1}{c}{(5)} & \multicolumn{1}{c}{(6)} & 
\multicolumn{1}{c}{(7)} & \multicolumn{1}{c}{(8)} & \multicolumn{2}{c}{(9)} & 
\multicolumn{2}{c}{(10)} \\
\multicolumn{1}{c}{date} & \multicolumn{1}{c}{JD} & telescope & instrument 
 & activity & spectral range & \multicolumn{1}{c}{res.} 
 & \multicolumn{1}{c}{no.} & \multicolumn{2}{c}{dur.} & \multicolumn{2}{c}{exp.}\\
\noalign{\smallskip}
\hline
\noalign{\smallskip}
Jan 1991 & 2\,448\,270 & ROSAT XRT & PSPC & RASS & 0.1-2.4 keV
 & & 1\hspace{1mm} & 1.5 & days & 8 & min \\
Aug 1991 & 2\,448\,500 & ESO/MPI 2.2m & EFOSC 2 & spec. & 3500-9000 \AA 
 & 40 \AA\hspace{1mm} & 1\hspace{1mm} & 10 & min & 10 & min \\
Jan 1992 & 2\,448\,632 & ESO/MPI 2.2m & EFOSC 2 & spec. & 3500-9000 \AA 
 & 40 \AA\hspace{1mm} & 9\hspace{1mm} & 4.1 & days & 5 & min \\
Jan 1992 & 2\,448\,633 & ESO/MPI 2.2m & EFOSC 2 & spec. & 3500-5400 \AA 
 & 10 \AA\hspace{1mm} & 23\hspace{1mm} & 1.1 & days & 10 & min \\
Jul 1992 & 2\,448\,832 & ROSAT XRT & PSPC & point. & 0.1-2.4 keV
 & & 0\hspace{1mm} & 2.4 & days & 3.8 & hours \\
Aug 1992 & 2\,448\,859 & ESO/MPI 2.2m & EFOSC 2 & spec. & 3500-5400 \AA 
 &  9 \AA\hspace{1mm} & 22\hspace{1mm} & 6.0 & days & 10 & min \\
Aug 1992 & 2\,448\,859 & ESO/MPI 2.2m & EFOSC 2 & spec. & 5800-8400 \AA 
 & 12 \AA\hspace{1mm} & 21\hspace{1mm} & 6.0 & days & 5 & min \\
Feb 1993 & 2\,449\,035 & ESO 1.0m & Photom. & phot. & UBVRI 
 & & 12\hspace{1mm} & 2.0 & days & 2 & min \\
Feb 1993 & 2\,449\,037 & ESO/MPI 2.2m & EFOSC 2 & spec. & 3500-9000 \AA
 & 25 \AA\hspace{1mm} & 1\hspace{1mm} & 5 & min & 5 & min \\
Feb 1993 & 2\,449\,038 & ESO/MPI 2.2m & EFOSC 2 & phot. & V 
 & & 10\hspace{1mm} & 2.0 & days & 30 & sec \\
Feb 1993 & 2\,449\,038 & ESO/MPI 2.2m & EFOSC 2 & spec. & 5800-8400 \AA
 & 8 \AA\hspace{1mm} & 27\hspace{1mm} & 2.1 & days & 10 & min \\
Aug 1993 & 2\,449\,224 & ESO/MPI 2.2m & EFOSC 2 & spec. & 3500-9000 \AA 
 & 25 \AA\hspace{1mm} & 1\hspace{1mm} & 15 & min & 15 & min \\
Jul 1994 & 2\,449\,566 & ROSAT XRT & HRI & point. & 0.1-2.4 keV
 & & 0\hspace{1mm} & 3.3 & days & 4.9 & hours \\
Feb 1995 & 2\,449\,752 & ESO 1.0m & Photom. & phot. & BVR 
 & & 102\hspace{1mm} & 4.2 & hours & 2 & min \\
Jul 1995 & 2\,449\,902 & ESO/MPI 2.2m & EFOSC 2 & spec. & 3500-5400 \AA 
 & 8 \AA\hspace{1mm} & 22\hspace{1mm} & 1.1 & days & 10 & min \\
Jul 1995 & 2\,449\,902 & ESO/MPI 2.2m & EFOSC 2 & spec. & 3800-9100 \AA 
 & 35 \AA\hspace{1mm} & 6\hspace{1mm} & 2.0 & days & 5 & min \\
Jul 1995 & 2\,449\,907 & ESO 1.54m & Dir. Im. & phot. & BVR 
 & & 213\hspace{1mm} & 1.1 & days & 1 & min \\
Dec 1995 & 2\,450\,075 & ESO/MPI 2.2m & EFOSC 2 & spec. & 3800-9100 \AA 
 & 30 \AA\hspace{1mm} & 14\hspace{1mm} & 2.0 & days & 10 & min \\
Jan 1996 & 2\,450\,097 & ESO/Dutch 0.9m & Dir. Im. & phot. & VRI 
 & & 242\hspace{1mm} & 4.2 & days & 30 & sec \\
Mar 1997 & 2\,450\,511 & ESO/MPI 2.2m & EFOSC 2 & spec. & 3600-10200 \AA 
 & 50 \AA\hspace{1mm} & 39\hspace{1mm} & 2.2 & days & 10 & min \\
Mar 1997 & 2\,450\,511 & ESO/MPI 2.2m & EFOSC 2 & spec. & 6440-8360 \AA 
 & 5 \AA\hspace{1mm} & 15\hspace{1mm} & 4.2 & hours & 15 & min \\
Mar 1997 & 2\,450\,513 & ESO/MPI 2.2m & EFOSC 2 & spec. & 3600-5200 \AA 
 & 6 \AA\hspace{1mm} & 16\hspace{1mm} & 4.6 & hours & 15 & min \\
May 1998 & 2\,450\,935 & ESO 3.6m & EFOSC 2 & phot. & BVR 
 & & 37\hspace{1mm} & 4.6 & hours & 1 & min \\
May 1998 & 2\,450\,937 & ESO 3.6m & EFOSC 2 & sp.pol. & 3600-7490  \AA 
 & 5 \AA\hspace{1mm} & 9\hspace{1mm} & 2.7 & hours & 15 & min \\
\noalign{\smallskip}
\hline
\end {tabular}
\end {flushleft}
\end {table*}

Polars or AM Herculis binaries belong to the class of cataclysmic
variables. They consist of a low mass main sequence star filling
its critical Roche lobe and a magnetic white dwarf accreting matter
from its companion. The strong magnetic field channels the accretion flow
into a small area on the surface of the white dwarf. In this area the
temperature rises to values of several 100\,000 K and shifts
the emission of reprocessed radiation into the soft X-ray regime.
Therefore an efficient way to detect these systems is to search the
sky for (highly) variable extreme ultraviolet or soft X-ray emitters.
The great success of ROSAT, ASCA, and EUVE in detecting these systems is
demonstrated by the fact, that before ROSAT only 17 Polars were
known (Cropper~1990) whilst as of today this number has increased to 63.
Other sources of radiation in these systems are thermal bremsstrahlung
emission from the accretion column, observed at higher
X-ray energies (typically 10 to 20~keV), cyclotron emission from the
accretion column due to the strong magnetic field, often dominating the
optical regime, and radiation from the secondary star, strongest
in the infrared and sometimes only being detectable, if the system is
in a state of low accretion. Detailed discussions of these systems
can be found in the books by Warner (1995) and Campbell (1997).
Early reviews on the basis of the new ROSAT data have been presented among
others by Beuermann \& Thomas (1993), Watson (1994), Beuermann \& Burwitz (1995),
and Schwo\-pe (1995), a more recent one by Beuermann (1997). The distribution
of Polars in the solar neighborhood was investigated by Thomas \& Beuermann (1997).
An identification program based on a complete sample of
the brightest soft X-ray sources from the RASS (Voges et al.~1999) at high
galactic latitudes (Thomas et al.~1998) led to the detection of \RX{}.

\section{Observations} 

Starting 12 January 1991 the position of \RX{} in the sky was scanned by
the ROSAT XRT with the PSPC as detector. During 20 satellite orbits with
a total exposure time of 485 s the object was detected with a mean count rate
of 1.8 \cts\ and a hardness ratio $HR1 = -0.85$\footnote{$HR1 = (H-S)/(H+S)$ with
$H$ and $S$ the count rates in the hard and soft energy intervals
$0.5-2.4$\,keV and $0.1-0.4$\,keV, respectively.}. On a finding chart produced
from the COSMOS scans of the SERC-J plates at ROE (Yentis et al.~1992)
a V $= 16^{\rm m}$ star was found to be the likely optical counterpart,
at a distance of $\simeq$~3\arcsec\ from the X-ray position (Fig.~1).
A low resolution spectrum taken on 31 August 1991 at the ESO/MPI 2.2m telescope
showed strong Balmer and helium line emission thus confirming
this identification. A list of all observations collected until now can be found
in Table~1.

\subsection{X-ray photometry and spectroscopy}

From the RASS data a photon event table was extracted which covered
$50 \times 50$ arcmin$^2$ around the X-ray position of the source.
Using the EXSAS software package provided by the MPE Garching (Zimmermann
et al.~1994) for the extraction of source photons in a circle of 250\arcsec\
radius centered on the source and background photons in a circle of 400\arcsec\
radius in scan direction we obtained the light curve shown in Fig.~2 and the mean
spectrum inserted in Fig.~4. Phasing of the light curve has been obtained
from Eq.~(2) based on optical observations (see Sect.~3.2), so phase zero
corresponds to the inferior conjunction of the secondary star. The error in the
period results in a phase error below 0.01, therefore the cycle count is correct.
The data are sampled from nine different orbital cycles
of the binary. The mean count rate obtained from the light curve is 1.94 \cts.
In one ROSAT orbit a count rate far above the average (6.5 \cts) was measured. 
We consider this as a singular event not typical for the orbital variation.
Without this point the mean count rate drops to 1.68 \cts. In order to show
that the light curve is not dominated by changes between binary orbits we have
used two different symbols in Fig.~2 for the first and second half of the observation.
A period search without the point at 6.5 \cts\ revealed a likely period of
$254 \pm 7$~min.
It should be noted that at no phase did the count rate drop to zero, so the
accretion area giving rise to the X-ray emission never completely vanishes from view
behind the horizon of the white dwarf. The alternative of a second accretion area
contributing to the X-ray flux seems to be ruled out by the results of our
spectropolarimetry, which shows no change of sign in the circular flux (see Sect.~2.3).

\begin{figure}[t]
 \centerline{
 \epsfxsize=8.8cm
 \epsfbox[75 143 510 748]{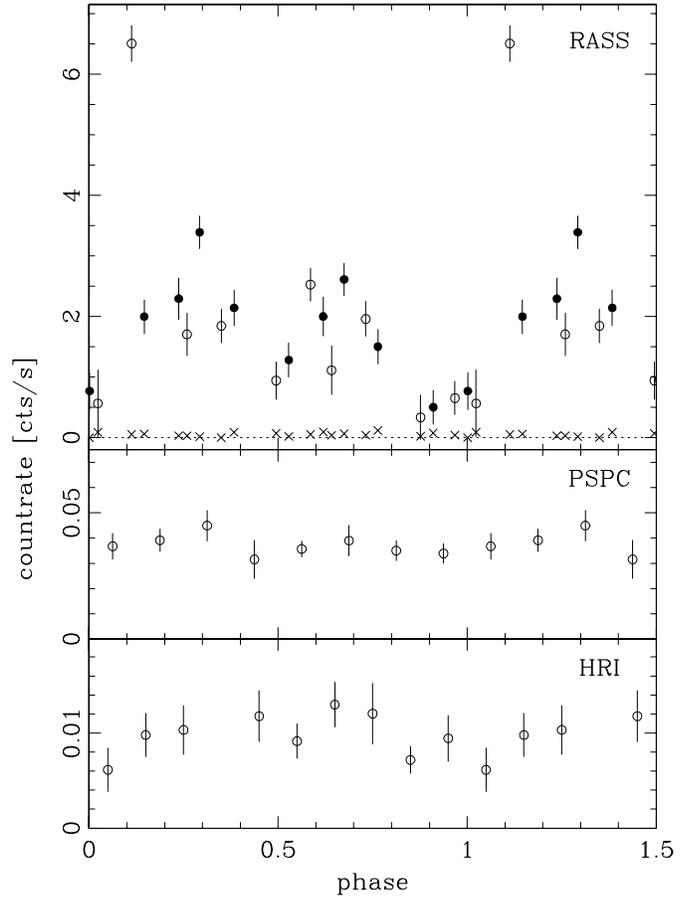}}
 \caption{X-ray light curve of \RX{}. The upper panel shows the source
 during the RASS (first half of observation: filled circles, second half:
 open circles) together with the measured background (crosses),
 the lower panels during two subsequent pointings with
 the PSPC and the HRI as detectors, respectively. For the preliminary period
 determination the single high datum at 6.5 \cts\ has been ignored. The phases
 are computed from Eq.~(2) in Sect.~3.2. }
\end{figure}

In two subsequent pointings with the ROSAT PSPC (July 1992) and HRI (July 1994)
as detectors the source was observed with count rates of 0.036~PSPC~\cts\
(hardness ratio $HR1 = 0.28$) and
0.010~HRI~\cts\ (corresponding to $\simeq$~0.06~PSPC~\cts). This is a reduction
in the count rate by factors 50 (1992) and 30 (1994) compared to the RASS.
Phase binning of these data resulted in the lightcurves shown in Fig.~2.
Again we note that the site of the X-ray emission always remains in view of
the observer.

\begin{figure}[t]
 \includegraphics[width=8.8cm,viewport=10 0 757 525,clip=]{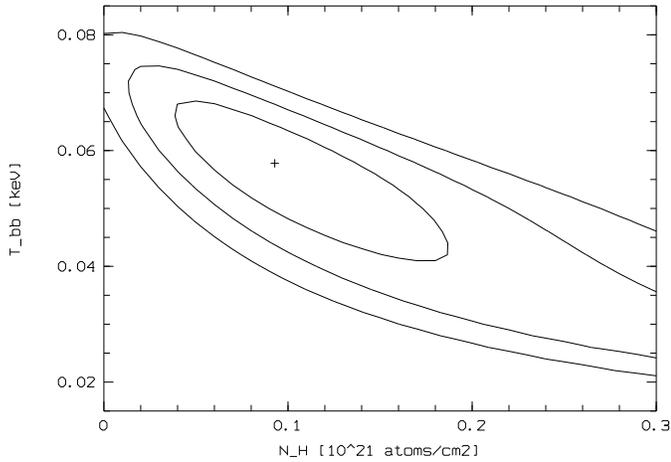}
 \caption{Confidence levels for the X-ray spectral fit of \RX{}. The 1, 
 2, and 3$\sigma$ ranges for the fit parameters blackbody temperature and
 column density are shown together with the fit result (cross).
 }
\end{figure}

Fitting the RASS spectrum with a blackbody and
a thermal brems\-strah\-lung component together with possible interstellar
absorption resulted in a blackbody temperature of 58\,eV and a
column density of $9\cdot 10^{19}$ \mbox{H-atoms\,cm$^{-2}$}. 
The uncertainties for these parameters are
depicted in Fig.~3. The (unabsorbed) contributions of the two
components to the total flux in the ROSAT window (0.1 to 2.4~keV)
amount to $1.14\cdot 10^{-11}$\,erg\,cm$^{-2}$s$^{-1}$ and
$0.07\cdot 10^{-11}$\,erg\,cm$^{-2}$s$^{-1}$, respectively. For the temperature
of the thermal brems\-strah\-lung component we assumed a value of 10\,keV,
since this cannot be derived from the spectral data. The resulting unabsorbed
spectra are shown in Fig.~4 as dotted lines. Separating the count rate into the
two components one obtains 1.70 \cts\ for the blackbody component and
0.07 \cts\ for the thermal brems\-strah\-lung component. 
The flux ratio $F_{\rm brems}/F_{\rm bb}$ in the ROSAT band is 0.06, integrated
over all frequencies it increases to 0.16.
The uncertainty in these values is about a factor 2.

\begin{figure}[ht]
 \centerline{
 \epsfxsize=8.8cm
 \epsfbox[115 69 722 515]{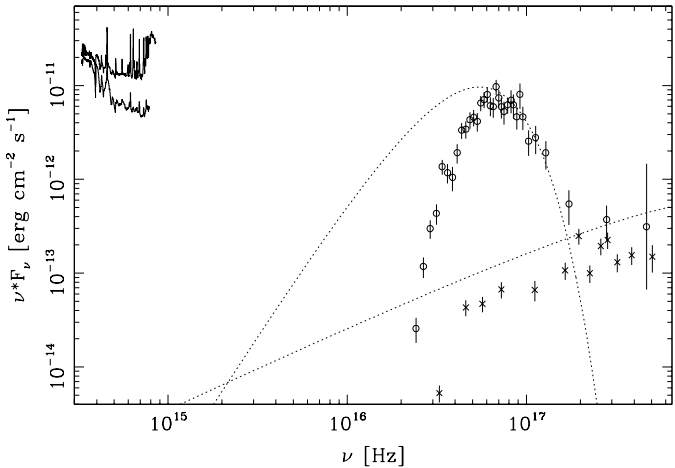}}
 \caption{Overall observed spectrum of \RX{}. The optical spectra shown at
 low frequencies were obtained in Feb. 93 (high state) and in Dec. 95 (low state).
 In the X-ray regime the RASS data are shown as circles, the PSPC pointing data
 as crosses. The dotted lines give the unabsorbed blackbody and bremsstrahlung
 components of the X-ray spectral fit to the RASS data.
 }
\end{figure}

The spectral fit to the data from the PSPC pointing gave a blackbody
temperature of 50\,eV, an unabsorbed flux of $4.7\cdot 10^{-14}$\,erg\,cm$^{-2}$s$^{-1}$
for the blackbody component and $9.1\cdot 10^{-13}$\,erg\,cm$^{-2}$s$^{-1}$ for the
thermal brems\-strah\-lung component, with large uncertainties and a high correlation
between flux and temperature of the blackbody fit (see Sect.~3.4). Here we again fixed
the temperature for the thermal brems\-strah\-lung component
to 10\,keV, and the column density to the value obtained during the RASS.
The data are also shown in Fig.~4.

\subsection{Optical photometry} 

Sequences of direct images in different filters were taken at several
epochs (see Table~1). To demonstrate the long-term variability of \RX{}
we have plotted in Fig.~5 the mean values of $V$ in the different runs together
with the range of variability defined by the magnitudes which bracket 90\%
of the measurements in the corresponding observing run. Added to this plot
are V-magnitudes from spectroscopy, obtained by folding the mean spectrum of
each observing run with the sensitivity of the V-filter
(a flux of $3.64\cdot 10^{-9}$\,erg\,cm$^{-2}$s$^{-1}$\AA$^{-1}$
corresponds to zero magnitude, see Bessell~1979). The dotted
lines indicate the times of the X-ray observations (RASS, PSPC pointing, and
HRI pointing). Typically \RX{} is found at $V \simeq 16^{\rm m}$ and brightened only
for short episodes near HJD 2\,449\,036 (Feb.\,93) and HJD 2\,450\,936 (May\,98).

The orbital variation of \RX{} in the V-band is displayed in Fig.~6.
Different symbols identify the different observing runs which are listed at
the bottom of the figure. The lightcurves are clearly of different character
in the bright (Feb. 93, May 98) and faint (all other) states of \RX{}.
While the bright states show only one minimum per orbit, the variation in the
faintest state (Feb. 95) almost follows a sinusoidal variation with two minima
per orbit. With increasing brightness the flux first increases outside the primary
minimum (near phase zero) and starts to deviate from its value around the primary
minimum for the bright states only.

It is worth mentioning that the observations in Feb.~93 were collected in six
subsequent nights with two different instruments and
that during this high state episode the orbital-averaged flux decreased by 
0.07\,mag per day, suggesting a characteristic duration of the high state of
a few weeks. The orbital light curve for the Feb.\,93 data in Fig.~6 (diamonds)
was obtained by subtracting a linear trend from the data. The fitted light curve
(dotted line) represents a sinusoidal fit to the detrended data.

\subsection{Optical spectroscopy and spectropolarimetry} 

Time sequences of spectra at different resolutions were obtained at 9 epochs as listed
in Table~1 and displayed in Fig.~5 (stars). We first display in Fig.~7 the low
resolution spectra from Feb.~93 (brightest state) and from July~95 (faintest state).
Both spectra clearly show the contribution of a late-type star to the flux at the
red wavelengths, identifiable through its TiO absorption troughs. The emission
lines of hydrogen and helium are visible in both spectra, with \Ion{He}{II}\,$\lambda$4686
strongly reduced in the fainter spectrum.

\begin{figure}[t]
 \centerline{
 \epsfxsize=8.8cm
 \epsfbox[119 70 743 510]{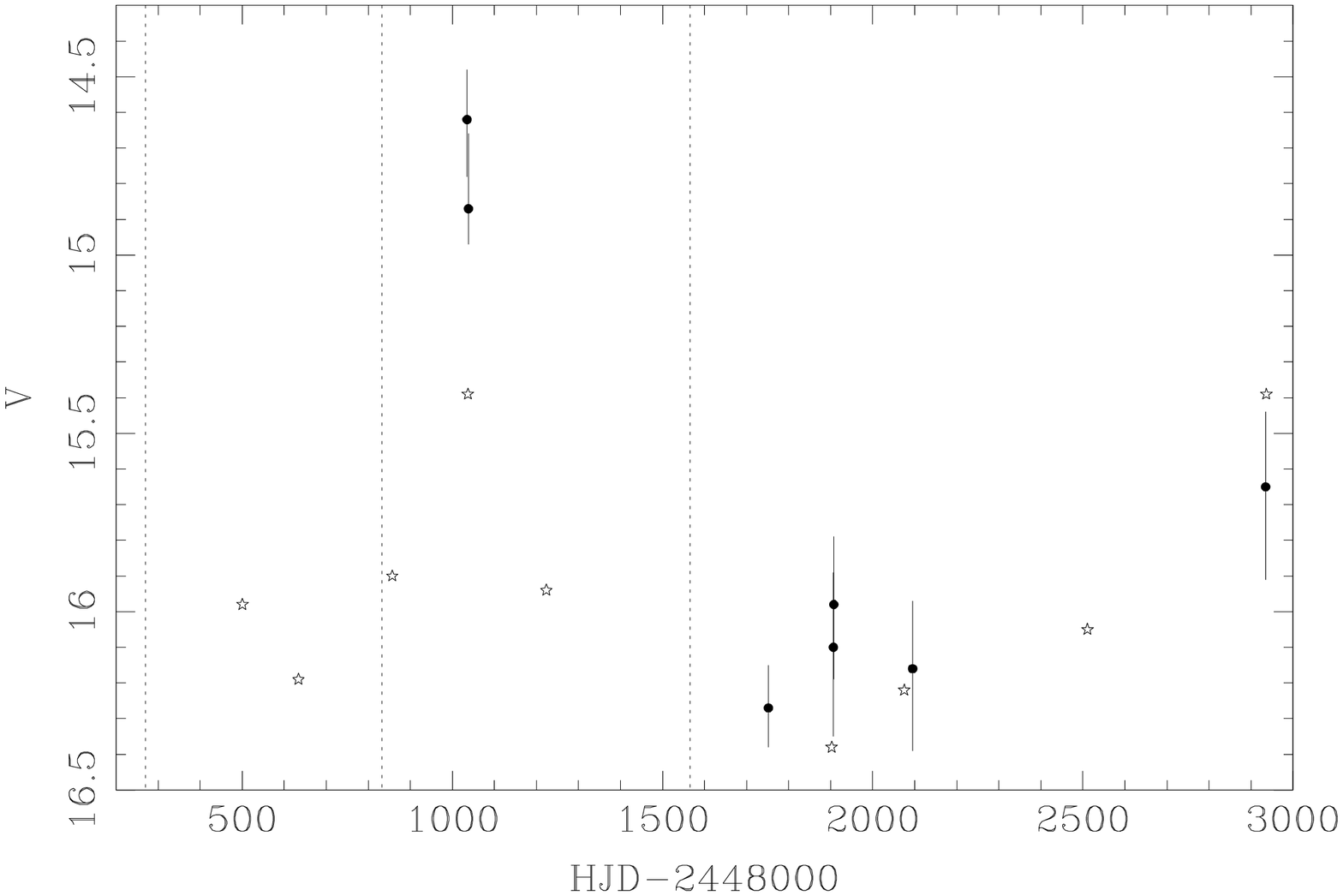}}
 \caption{Long-term light curve of \RX{}. Photometric observations are
 marked with filled dots (mean values) and a vertical bar (90\% range),
 magnitudes derived from mean spectra (less accurate) by stars. The dotted
 lines show the times of X-ray observations (RASS, PSPC, HRI).}
\end{figure}

\begin {table}[ht]
\caption[]{ Results from double Gaussian fits. The columns list the fitted emission line,
the velocity amplitudes in km/s and phases for blue-to-red zero crossing both for the
narrow and the broad component and the line flux ratio of minimum to maximum for the narrow
component.} 
\begin {flushleft}
\begin {tabular}{lccccc}
\noalign{\smallskip}
\hline
\noalign{\smallskip}
line & $v_{\rm narrow}$ & $\phi_0$ & $v_{\rm broad}$ & $\phi_0$ & flux ratio \\
\noalign{\smallskip}
\hline
\noalign{\smallskip}
H$\alpha$                    & 100 & 1.011 & 397 & 0.717 & 0.41 \\
H$\beta$                     & 109 & 0.994 & 353 & 0.709 & 0.31 \\
H$\gamma$                    &  97 & 0.976 & 485 & 0.687 & 0.29 \\
H$\delta$                    & 106 & 0.992 & 434 & 0.704 & 0.27 \\
\Ion{He}{I}\,$\lambda$4026   & 101 & 0.933 & 624 & 0.673 & 0.22 \\
\Ion{He}{I}\,$\lambda$6678   &  92 & 0.957 & 498 & 0.714 & 0.24 \\
\Ion{He}{II}\,$\lambda$4686  &  63 & 0.937 & 512 & 0.712 & 0.26 \\
\Ion{Ca}{II}~K               & 117 & 0.972 & 567 & 0.714 & 0.24 \\
\noalign{\smallskip}
\hline
\end {tabular}
\end {flushleft}
\end {table}

\begin{figure*}[t]
 \centerline{
 \epsfxsize=8.8cm
 \epsfbox[252 83 574 509]{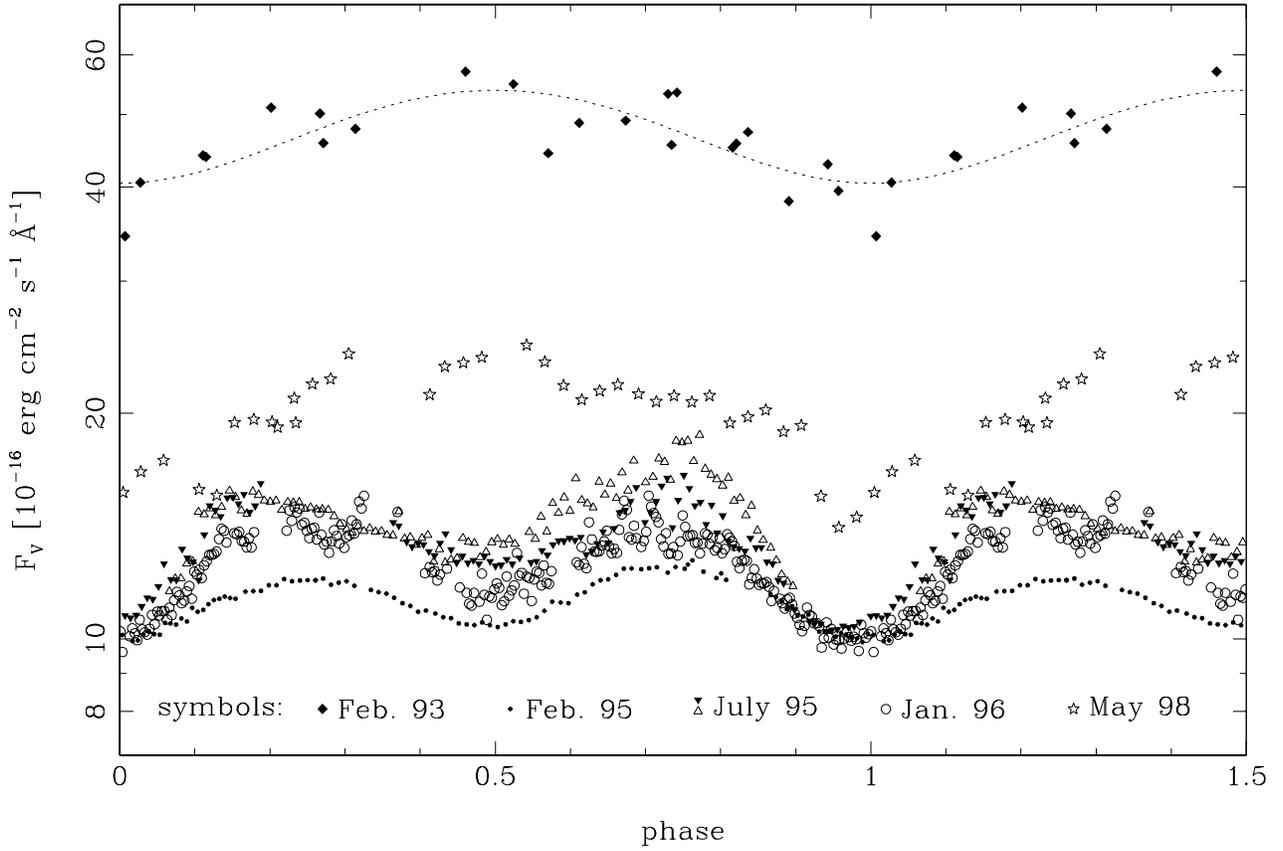}}
 \caption{Phase-folded lightcurves at different observing times as given
  at the bottom of the plot. For the dotted line see text in Sect.~2.2.
 The phases are computed from Eq.~(2) in Sect.~3.2.}
\end{figure*}

Complete coverage of the orbit in one night could be achieved for the observation
run in March~97 (intermediate state) only. Medium resolution spectra
in the blue (3600--5200~\AA) and
in the red (6440--8360~\AA) were analyzed for Doppler shifts in their emission
and absorption lines. The most accurate results were obtained for the red spectra.
Fitting the H$\alpha$
emission line with a double Gaussian profile splits the line into a narrow
(average FWHM 5.9~\AA, unresolved) and a broader (average FWHM 17~\AA) component. The resulting
radial velocities are plotted in Fig.~8, upper panel. The average radial velocities of three
absorption lines (\Ion{Na}{I}\,$\lambda$8183, 8195 and \Ion{K}{I}\,$\lambda$7699) are shown
in the same panel.
In the lower panels of Fig.~8, the fluxes in the
narrow emission component of \Ion{He}{I}\,$\lambda$6678 and the absorption line of
\Ion{K}{I}\,$\lambda$7699 are displayed together with results from irradiation computations
(see Sect.~3.2). The measured fluxes of the absorption line are very sensitive to
the assumed level of the continuum near this line and may be in error by up to 30\%.
We also measured the fluxes in the narrow emission components of
H$\alpha$, H$\beta$, H$\gamma$, H$\delta$, \Ion{He}{II}\,$\lambda$4686,
\Ion{He}{I}\,$\lambda$4026,
and \Ion{Ca}{II}~K by fitting double Gaussians to the profiles. In Table~2 the velocity
amplitudes and phases for the blue-to-red zero crossings are given for the two components
together with the line flux ratios (minimum to maximum) of the narrow components. The
phases were derived from the ephemeris given in Sect.~3.2. The minimum emission line flux
stays finite for all of these lines. In this respect, all lines behave similarly to
\Ion{He}{I}\,$\lambda$6678 in Fig.~8.

During the observation in Dec.~95 the source was in a low state, which allows for
a separation of the spectral flux into the different contributions from the M-star, the
accretion area, and the white dwarf. The results of this analysis will be discussed
in Sect.~3.3. The spectra taken during another low state in Jan.~92 are not suited
for this kind of analysis because they cover less than half an orbit and their flux
calibrations are unreliable.

For the observation in May~98 (high state) EFOSC~2 was equipped with the Wollaston prism and
a quarterwave plate, using Grism B300. The sequence of $2 \times 9$ spectra covers
only 60\% of the orbital period. Circularly polarized fluxes have been obtained by taking the
difference between the two spectra produced by the Wollaston prism (Fig.~9). All spectra show
circular polarization of negative sign only. We find, that two maxima of the polarized
flux, at $\simeq$\,5000\,\AA\ and $\simeq$\,6600\,\AA, are present in all spectra
except near phase 0 where the viewing angle is smallest with respect to the axis of the
accretion funnel. In addition, a third maximum of the polarized flux occurs at
$\simeq$\,3950\,\AA\
over the restricted phase interval of 0.3 to 0.5. We will argue below that these are
the $3^{\rm rd}$, $4^{\rm th}$, and $5^{\rm th}$ harmonic in a field of about 56\,MG and
that the different phase behavior arises from optical depth and geometric effects (Sect.~3.3).

\begin{figure}[t]
 \includegraphics[width=8.8cm,viewport=99 64 702 500,clip=]{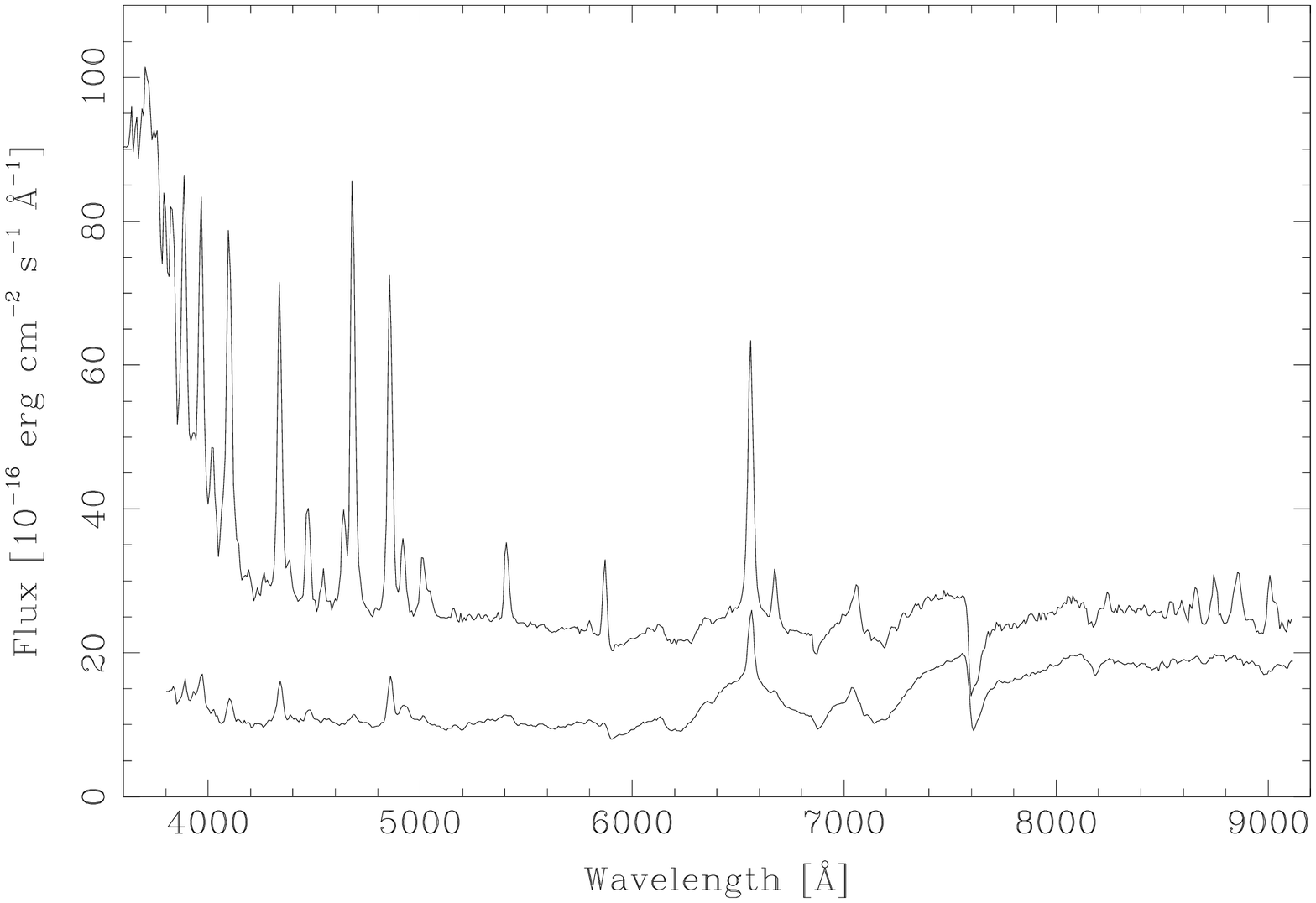}
 \caption{Spectra at low resolution from two different observing runs. 
 The brighter spectrum was taken in Feb. 93, the other is the mean of six
 spectra obtained in July 95.}
\end{figure}

Spectra taken during the other observing runs have been analyzed for radial velocities,
flux contribution from the secondary star, and cyclotron emission. Because of lower
resolution, incomplete orbital coverage, or less favorable observing conditions they
mainly helped to reinforce and confirm the results obtained from the March~97 data and
are important for excluding possible alias periods.

\begin{figure}[t]
 \centerline{
 \epsfxsize=8.8cm
 \epsfbox[77 134 536 713]{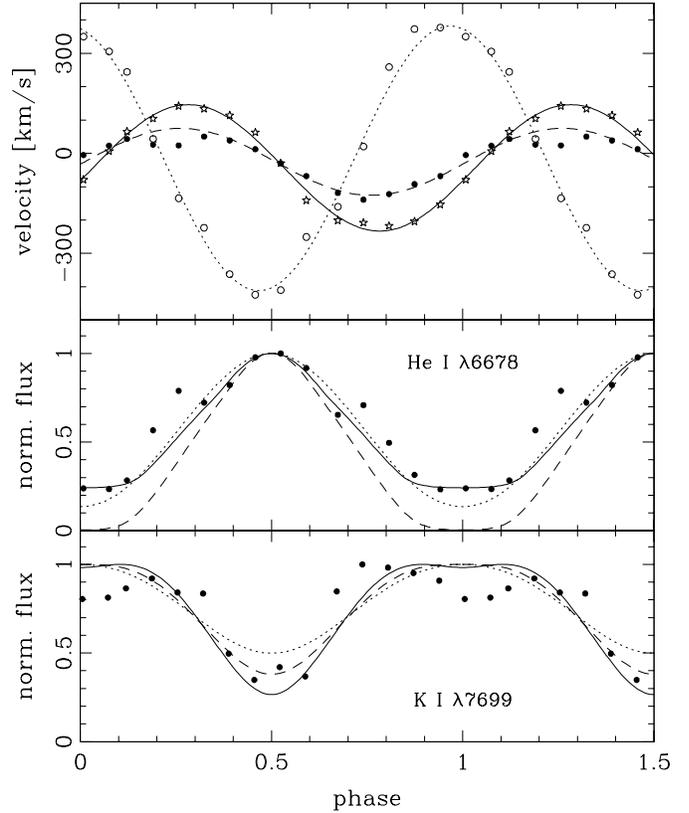}}
 \caption{Line properties extracted from spectra taken in Mar.~97. {\it Upper panel:}
 radial velocities averaged from measurements of three absorption lines (stars), from
 the narrow component of the H$\alpha$ emission line (filled circles), and the broad
 component of the H$\alpha$ emission line (open circles). The solid, dashed, and dotted
 lines give the sinusoidal fits to the data. {\it Middle panel:} normalized line flux of the
 \Ion{He}{I}\,$\lambda$6678 narrow emission component together with results from irradiation
 computations for three different inclinations (solid: $66^\circ$, dashed: $52^\circ$,
 dotted: $31^\circ$, see Sect.~3.2).
 {\it Lower panel:} normalized line flux of the \Ion{K}{I}\,$\lambda$7699 absorption line
 together with results from irradiation computations for the same inclinations.
 }
\end{figure}

\section{Results} 

\subsection{A distance estimate}

In all low resolution spectra the TiO absorption bands from the secondary are
easily recognized. We therefore use the calibration method decsribed in
Beuermann \& Weichhold (1999)
to determine the surface flux from the flux difference between 7165 and
7500~\AA. In July and Dec. 95 \RX{} was in a low state, so we used the flux
differences as observed. For the Mar. 97 data we first estimated the
gradient of the cyclotron emission between 7165~\AA\ and 7500~\AA\ by subtracting
suitably scaled spectra of the M-dwarf Gl~207.1 and corrected the measured flux
differences for that. We also tried to fit spectra of other M-dwarfs (e.g. Gl 205,
Gl 352) but with less satisfactory results.
From these 35 spectra we obtained a mean value of
$9.7 \pm 1.1\cdot$\fluxunit{} for the flux difference.
This value is reduced by 2\% if
we exclude all spectra with phases between 0.25 and 0.75, i.e. looking at the unilluminated
backside of the secondary. In the following we use a value of $9.5\cdot$\fluxunit{}.

\begin{figure}[t]
 \epsfxsize=8.8cm
 \epsfbox[70 524 522 760]{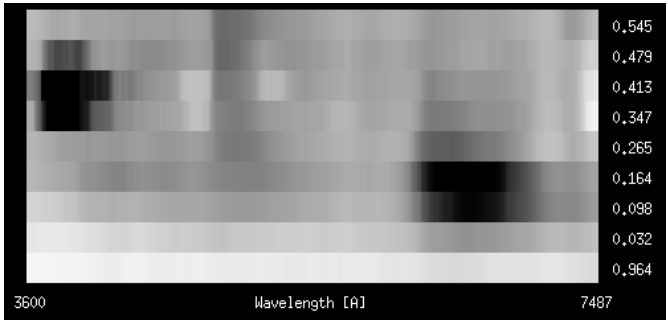}
 \caption{Circularly polarized flux extracted from spectra taken in May~98. 
 The grayscale plot displays fluxes between 0 (white) and
 $-6\cdot$\fluxunit{} (black),
 orbital phases are given on the right-hand side. The image is smoothed in order to
 remove the contributions from the emission lines.
 }
\end{figure}

Filling its critical Roche lobe the secondary must have a mean density of
$\rho/\rho_\odot = 4.46$ for a system period of 4.19\,h. This value depends slightly on the
mass ratio but the distance changes by no more than 2\% for white dwarf masses between
0.5 and 1.2~M$_\odot$. Assuming the secondary to be a main-sequence star the stellar
models of Baraffe et al. (1998) suggest a mass of 0.51~M$_\odot$ and a radius of
0.486~R$_\odot$, with a slight dependence on mass ratio. However, the secondary, due to its
mass-loss history, may be out of thermal equilibrium and therefore have a lower mean density
than a main-sequence star of the same mass. Comparing the low-state spectra to those of
several M-stars we
estimate the spectral type to be M2.5 $\pm 0.5$. Since the spectral type does not change
drastically for stars out of thermal equilibrium (Kolb \& Baraffe 1999), we obtain
an estimate for the
mass of the secondary from a comparison with models of Baraffe et al. (1998):
$M_2/{\rm M}_\odot \simeq 0.45$. Using the relation between spectral type and the surface
brightness F$_{\rm TiO}$ by Beuermann \& Weichhold (1999) we arrive at a distance of

\begin{equation}
200 \pm 17\ {\rm pc}\cdot (M_2/0.45\,{\rm M}_\odot)^{1/3}.
\end{equation}

\subsection{System parameters}

The orbital period of \RX{} was obtained from a sinusoidal fit to the radial
velocity measurements of the absorption lines \Ion{Na}{I}\,$\lambda$8183,8195 and
\Ion{K}{I}\,$\lambda$7699,
obtained in Aug.~92, Feb.~93, and Mar.~97. Possible aliases have been checked against
other radial velocity measurements obtained at many different epochs (see Table~1). The
zero point for the resulting ephemeris is defined as the blue-to-red zero-crossing of
the radial velocity curves. The fits to the narrow emission line components of the
Mar.~97 data give a slightly earlier zero point than the absorption line data
(difference in phase: $0.06 \pm 0.03$) and we have used the mean of both data sets.
The resulting ephemeris is (errors of the last digits are given in brackets):

\begin{equation}
T = {\rm HJD}\ 2\,448\,632.4435 (43) + 0.174\,592\,09 (15)\ E
\end{equation}

\noindent
The relatively large error in the zero point reflects the difference in determining the
blue-to-red zero-crossing from both the narrow emission and the absorption line measurements.
The total time span between our first and last observation amounts to 2667 days or 15276 cycles.
Thus the maximum error in the phasing calculated from the error for the period is 0.013,
which excludes any errors in the cycle count.

The measurements shown in the two lower panels of Fig.~8 clearly demonstrate that the
narrow emission line flux is stronger and the absorption line flux weaker than average when
the illuminated side of the secondary is in view of the observer. Such a behavior can be
understood with the irradiation model of Beuermann \& Thomas (1990). Using the amplitude
of $185 \pm 9$ km/s measured for the absorption lines the model determines the inclination
angle as a function of mass ratio. We have assumed that the contribution to the absorption
line flux from the illuminated side is negligible.
For primary masses of 0.5 and 1.1\,M$_\odot$ we obtain
inclinations of 52\degr\ and 31\degr, respectively, taking 0.45\,M$_\odot$ for the mass of
the secondary. The corresponding velocities for the emission lines are then 51 and 98~km/s.
Looking at the measured values (Table~2) this would favor the low inclination and so a high
primary mass. A low inclination also roughly fits the observed line fluxes, an example is
shown in Fig.~8
(middle panel, dotted line). For the inclination of 66\degr\ derived below (Sect.~3.3) this
is not the case. But if we assume that a substantial contribution to the line flux comes
from the unilluminated side of the secondary the situation changes. To reproduce the observed
average amplitude of 102.6 km/s for the narrow emission lines, $\simeq 55$\% of the line flux
(averaged over the surface of the secondary) must be generated by illumination. We then
obtain the line flux variation displayed in Fig.~8 (middle panel, solid line), which fits the
observed variation best, except for those measurements where the radial velocity curves of
broad and narrow component cross each other and therefore a separation into the two
components is difficult to achieve. While this result depends only weakly on the
assumed inclination it shows
that the observed line flux and velocity amplitude for the narrow emission are consistent with
a high inclination if part of the emission line flux is generated on the unilluminated side
of the secondary.

\begin{figure}[t]
 \includegraphics[width=8.8cm,viewport=28 16 517 748,clip=]{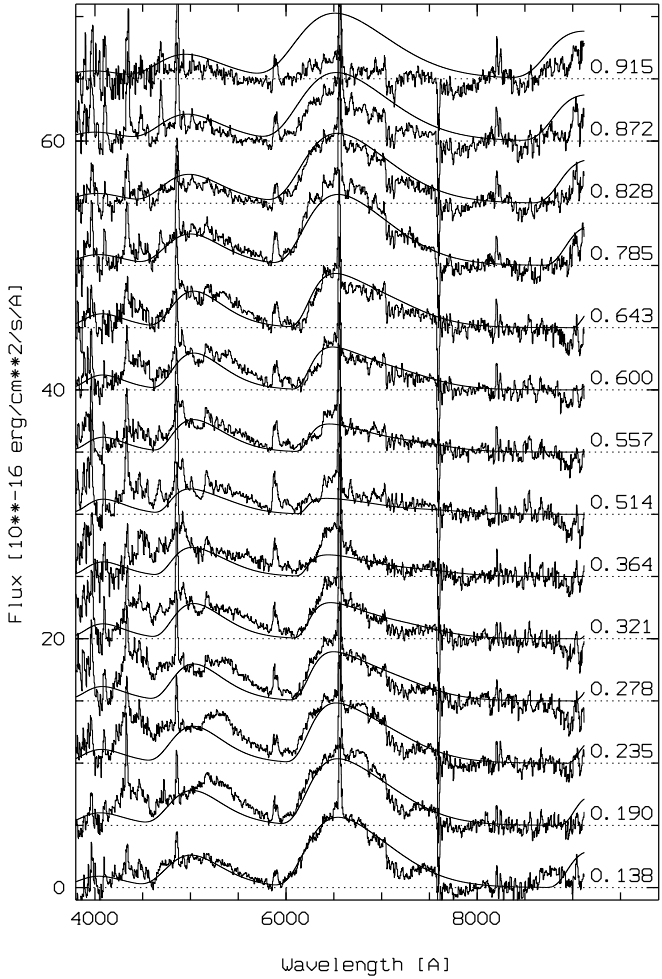}
 \caption{Cyclotron flux extracted from spectra taken in Dec~95. 
 The model fits are discussed in Sect.~3.3, orbital phases are given on the right-hand side.
 }
\end{figure}

\subsection{Cyclotron emission}

For the data taken during a low state in Dec.~95, we represent the
spectral flux in the 14 individual low-resolution spectra by the sum of a Rayleigh-Jeans
spectrum,
an M-star spectrum, and a cyclotron spectrum. This was done by adjusting the Rayleigh-Jeans
to the flux between 4040 and 4080~\AA\ and the M-star template to the flux
difference between 7165 and 7500~\AA. (As discussed in Sect.~3.1
the best M-star template is that of the M3 star Gl~207.1).
Subtracting these two contributions we are left with the cyclotron spectra shown in Fig.~10.
The variation of the Rayleigh-Jeans component in the V-band is roughly sinusoidal and
can be fitted with a mean of $3.1 \pm 0.1$ and an amplitude of $0.3 \pm 0.1$, both in
units of \fluxunit, the maximum occuring at phase $0.01 \pm 0.4$. Such a variation
indicates heating of an area around the accretion spot (see Sect.~3.3). 
The flux of the M-star displays maxima at phases near quadrature. The origin
of the these variations must be due to ellipsoidal
modulation from the M-star. The flux ratio between phases 0.5 (primary minimum) and
0.25/0.75 (maxima) amounts to $0.85 \pm 0.02$. 

To model the Dec.~95 variation of cyclotron flux with orbital phase we adapted the emission
of an isothermal homogeneous plasma slab (Barrett \& Chanmugam 1985) to the data.
The input parameters were the magnetic field strength $B$, the plasma
temperature $T$, the inclination $i$ of the system, the angle $\beta_f$ between the rotation
axis and the field direction, the angles $\beta_s$ and $\psi_s$ describing the location
of the accretion region (relative to the rotation axis and the line connecting the two stars),
and the dimensionless thickness $\Lambda$ and the area $A_{\rm s}$ of the accretion region. From
the phasing of the broad emission lines we took the azimuthal angle between field direction
and direction to the companion $\psi_f$ to be 17\degr. The fit shown in Fig.~10 required
$B = 56\,{\rm MG}$, $T = 10\,{\rm keV}$, $i = 66\degr$, $\beta_f$ = 7\degr,
$\beta_s$ = 24\degr,
$\psi_s$ = 18\degr, $\Lambda$ = 26, and $A_{\rm s} = 8.2\cdot 10^{16}$~cm$^2$ (for a
distance
of 200~pc). It was obtained in a iterative procedure starting with a coarse grid
of cyclotron spectra which allowed to fix $B$, $T$, $i$, and $\beta_f$ and then refining the
grid for the other parameters. The poor fit at phase 0.915 may be the result of
absorption by the accretion stream which is in front of the accretion area around phase
0.95. The rather high inclination can only be made consistent with the results
from the irradiation model for a white dwarf mass of 0.39\,M$_\odot$ (see Sect.~3.5). The
values of $\beta_f$ and $\beta_s$ are in conflict with the assumption of a pure dipolar
field configuration because
that requires $\beta_f > \beta_s$. Also this simple model does not reproduce the spectral
shape at short wavelengths, probably due to a more complicated accretion geometry than used here.

Using the parameters derived above we computed the gravity darkening in the Roche geometry.
For the ratio of primary minimum to maximum we obtained a value between 0.87
(without limb-darkening) and 0.83 (50\% limb-darkening), in agreement with the value
for the ellipsoidal variation deduced above.

For the values of $i$, $\beta_s$,
and $\psi_s$ the sinusoidal variation of the Rayleigh-Jeans component can be explained
assuming a constant contribution from the white dwarf photosphere plus a varying
contribution from the accretion area. The fluxes in the V-band then amount to 2.8 and
0.85, respectively, both in units of \fluxunit{}. At the derived distance
the constant flux gives an
absolute magnitude of M$_{\rm V} = 11.3^{\rm m}$, corresponding to a photospheric temperature
of $\simeq$\,14\,000\,K. The flux ratio of the variable to constant component
in an accretion area occupying a fraction $f$ on the white dwarfs surface requires a
temperature which is a factor $0.74~f^{-0.25}$ higher than the temperature of the white
dwarf, in reasonable aggreement with the results of G\"ansicke et al. (1999).

We also tried to fit the spectropolarimetric data from May~98, which do not suffer from the
uncertainties introduced by the subtraction of other flux contributions. But because of the
incomplete orbital coverage
we can only state that magnetic field strength, inclination, and field direction are
similar, while the thickness $\Lambda$ must be somewhat larger to produce the high circularly
polarized flux at 3950~\AA\ around phase 0.4 (see Fig.~9). We plan to repeat these observations,
which will then allow us to check the system parameters derived above. Cyclotron
spectra extracted from the Mar.~97 data show similar variations with phase as the data
of Dec.~95, but the removal of other flux contributions introduced too large an uncertainty,
so we did not try to fit cyclotron spectra to that dataset.

These considerations also allow us to qualitatively understand the behavior of the V-flux
variations (Fig.~6). In a low state both the cyclotron flux and the ellipsoidal variations
of the secondary cause a light curve with two maxima during one orbit, while in a high state
the contribution of the secondary is negligible and the cyclotron flux contribution most
likely comes from a more extended accretion area. The minimum around phase 0.5 which resulted
from a viewing angle close to 90\degr\ will then be filled in because of contributions from
other parts of the accretion area at lower viewing angles.

\subsection{X-ray emission}

We now turn to the X-ray light curves and spectra obtained during the RASS and the two
ROSAT PSPC and HRI pointed observations.
Since the inclination is larger than the field line direction it is expected that at
least part of the X-ray flux is absorbed by the stream of matter towards the white dwarf.
This explains the low flux around phase 0.95 both in the RASS and to some degree in
the HRI pointing. The reduction in the count rate by a factor 50 between the RASS and the
PSPC pointing and the corresponding change in the hardness ratio $HR1$ from $-0.96$ to 0.28
could partially be the result of a lower blackbody temperature. Although the best fit
temperature for the pointing is 50\,eV ($\chi^2$/d.o.f. 22/10), with a reduction of the blackbody
flux by a factor 230 compared to the RASS, the fit allows for much higher
blackbody fluxes if the temperature would be lower: at 15\,eV, the total blackbody flux
decreases by a factor 14 only compared to the RASS ($\chi^2$/d.o.f. 27/11). This
is of the same order as the reduction in cyclotron flux between the observations in Feb.~93
and Dec.~95 (factor 7.4). The contribution of the bremsstrahlung flux to the PSPC pointed
data is reduced by a factor 2.3 compared to the RASS.

\subsection{Masses of the components}

The masses determined above are in conflict with present understanding of stable mass
transfer. For donor stars with masses below $\simeq 0.5$\,M$_\odot$ mass transfer is
dynamically unstable for mass ratios M$_2$/M$_1 > 0.7$ (Webbink 1985). The lower
limit of 0.38\,M$_\odot$ for the mass of the secondary (Sect.~3.1) also does not fulfil
the criterion for stability. Even if we assume an evolved secondary as in Kolb
\& Baraffe (1999) we arrive at masses of 0.31\,M$_\odot$ for the white dwarf and
0.25\,M$_\odot$ for the M~star, again violating the stability criterion. The mass ratio
is determined from the observed
absorption line velocity and the inclination. So we turn the question around and ask,
for which inclination would the stability criterion be satisfied? Taking the derived
mass of the secondary we then need a primary mass of 0.64\,M$_\odot$ and an inclination
of 45.3\degr. That implies a variation of the viewing angle around this value, which
in turn causes cyclotron emission lines to be shifted to the blue as compared to our data
in Fig.~10. Especially above 8000\,\AA\ the spectral flux should strongly increase
with wavelength for all
phases due to the presence of the next lower harmonic, which in Fig.~10 only marginally shows
up around phase zero. Therefore with our data we can see no remedy to the present situation.
So the stability problem makes \RX{} an interesting system for future
observations in four aspects: a) the contribution of the secondary star to the spectrum,
best observable during a low state, b) the absorption line flux variation relevant for
modelling the illumination, c) the degree of circular polarization as a function of
orbital phase, and d) the white dwarf contribution to the spectrum, best observable
in the ultraviolet regime.

Finally, we note that constraining the mass of the white dwarf is important for our
understanding of the formation of cataclysmic variables, because with the period
determined above models of common envelope evolution predict that the mass of the
white dwarf should exceed 0.6\,M$_\odot$ (de Kool 1992).

\subsection{Mass accretion rates}

\begin {table}[ht]
\caption[]{Integrated fluxes from the accretion area and mass accretion rates.
The units are $10^{-11}$\,erg\,cm$^{-2}$s$^{-1}$ for the fluxes and
$10^{-11}$\,M$_\odot$\,yr$^{-1}$ for the mass accretion rates.
} 
\begin {flushleft}
\begin {tabular}{lcc}
\noalign{\smallskip}
\hline
\noalign{\smallskip}
 & high state & low state \\
\noalign{\smallskip}
\hline
\noalign{\smallskip}
cyclotron emission       & 1.7         & 0.2 \\
blackbody radiation      & 1.0 --- 3.0 & 0.0 --- 0.1 \\
thermal bremsstrahlung   & 0.1 --- 0.3 & 0.1 \\
\noalign{\smallskip}
\hline
\noalign{\smallskip}
total flux               & 2.8 --- 5.0 & 0.3 --- 0.4 \\
mass accretion rate      & 4.3 --- 7.7 & 0.5 --- 0.6 \\
\noalign{\smallskip}
\hline
\end {tabular}
\end {flushleft}
\end {table}

For an estimate of the mass accretion rate one has to sum up all flux contributions
from the accretion area. Although the optical and X-ray observations are not
simultaneous we take the optical observations in Feb.~93 and the X-ray observations
during the RASS as representative of the high state and those of Dec.~95 in the
optical and July~92 in X-rays of the low state. To extend the observations to the
whole frequency range we used the model fits for cyclotron emission, blackbody radiation,
and thermal bremsstrahlung, respectively. The main uncertainty in these values is caused
by the uncertainty in the blackbody temperature. Therefore we took the
$1\sigma$ upper and lower limits of 68 and 41~eV (see Fig.~3) for the RASS data and
assumed temperature limits of 15 and 50~eV for the PSPC data (see Sect.~3.4), to provide
a range for the derived fluxes. The results are summarized in Table 3. From
the total fluxes we computed the mass accretion rates for a distance of 200~pc, a
white dwarf mass of 0.4~M$_\odot$, and a white dwarf radius of $1.08\cdot 10^9$cm.
 
\section{Conclusions}

It has been shown that \RX{} belongs to the class of magnetic cataclysmic variables
called AM Herculis binaries or Polars. It has a period of 251.4~min, which places it
above the period gap, and a distance of $\simeq$~200~pc. From the occurrence
of cyclotron
humps a magnetic field strength of 56~MG for the main accreting pole has been deduced.
The inclination is $i \simeq$~66\degr\ implying a white dwarf mass of
$\simeq$~0.4~M$_\odot$, with a likely mass for the secondary of 0.45~M$_\odot$, corresponding
to a mass ratio $q \simeq 1.1$. 
A rough estimate of the mass accretion rate leads to $6\cdot 10^{-11}$\,M$_\odot$\,yr$^{-1}$
for the high state and $6\cdot 10^{-12}$\,M$_\odot$\,yr$^{-1}$ for the low state. The
rather low mass of the white dwarf required for reproducing the observed cyclotron
emission is in conflict with our present understanding of stable mass transfer and the
evolutionary history of cataclysmic variables, so further observations of this system
are of great interest.

\begin{acknowledgements}
The ROSAT project is supported by the Bundesministerium f\"ur Bildung und Forschung
(BMBF/DLR) and the Max-Planck-Gesellschaft. We thank the ROSAT team at the MPE (Garching)
for providing us with the results from the ROSAT All-Sky Survey and for their help with
the data analysis. We also thank the staff at La Silla for competent assistence during the
observations. This work has been supported in part by the DLR under grants 50\,OR\,9403\,5
and 50\,OR\,9210\,1.
\end{acknowledgements}

\end{document}